\documentclass[aps,pra,superscriptaddress,preprint]{revtex4-1}
\usepackage{subfigure}
\usepackage{xcolor}
\usepackage{graphicx}
\usepackage{bm}
\usepackage{amsmath}
\usepackage{float}
\usepackage{ulem}
\usepackage[utf8]{inputenc}
\usepackage[colorlinks,linkcolor=blue,anchorcolor=blue,citecolor=blue,urlcolor=blue]{hyperref}

\begin{document}

\title{Polarons in Binary Bose-Einstein Condensates}

\author{Ning Liu}
\email{corresponding author: ningliu@mail.bnu.edu.cn}
\affiliation{Department of Physics, Beijing Normal University, Beijing 100875, China}
\author{Z. C. Tu}
\affiliation{Department of Physics, Beijing Normal University, Beijing 100875, China}
\date{\today}
\begin{abstract}
Bose polarons are quasiparticles formed through the interaction between impurities and Bose-Einstein condensates. In this paper, we derive an effective Fr\"{o}hlich Hamiltonian using the generalized Bogoliubov transformation. The effective Fr\"{o}hlich Hamiltonian encompasses two types of effective interactions: impurity-density (ID) coupling and impurity-spin (IS) coupling. Furthermore, we employ the Lee-Low-Pines variational approach to investigate the relevant properties of Bose polarons induced by the ID and IS coupling. These properties include the ground state energy, effective mass, and average number of virtual phonons. Our findings reveal that the contribution resulting from IS couplings to the ground energy decreases to zero near the miscible-immiscible boundary. Additionally, the increase of the IS coupling induces a greater number of virtual phonons, impeding the movement of impurities and leading to a significant increase in the effective mass of Bose polarons.

\end{abstract}

\maketitle

\section{Introduction}
The concept of the polaron has intrigued many physicists since Landau's original idea of electron self-trapping in an ionic crystal~\cite{Landua1933}. However, systematically exploring the properties of polarons in solid materials has proven challenging~\cite{Skou2022}. Fortunately, advancements in experimental techniques have opened up new opportunities for observing polarons in ultracold atomic gases, potentially offering fresh insights into polaron properties. In particular, Bose polarons, which are quasiparticles formed by impurities immersed in a Bose-Einstein condensate (BEC), have been extensively studied in theoretical investigations~\cite{Grusdt2015, Huang2009, Novikov2009, Tempere2009, Ichmoukhamedov2019, Rath2013, Ardila2015, Ardila2019, Isaule2021, Shchadilova2016, Drescher2019, Guenther2018, Field2020, Volosniev2017, Khan2021}. Encouragingly, Bose polarons have been experimentally realized by two independent research groups~\cite{Hu2016, Jorgensen2016}, including cases near quantum criticality~\cite{Yan2020}. However, most existing studies on Bose polarons have been limited to single-component BECs, where the impurities interact solely with a single type of Bogoliubov excitation.

In the context of two-component BECs, the interaction between the condensates gives rise to two distinct Bogoliubov excitations, one associated with density fluctuations and the other with spin fluctuations~\cite{Pita2016, Abad2013, Recati2019, Ota2020, Cominotti2022}. Furthermore, the presence of phenomena such as phase separation~\cite{Pitaevskii2016} and quantum droplets~\cite{Petrov2015} has sparked interest in investigating the properties of polarons within two-component BECs. In recent years, several studies have focused on polarons in this context~\cite{Compagon2017, Charalambous2020, Boudjemaa2020, Keiler2021, Bighin2022}. However, despite these efforts, some fundamental questions remain unanswered. For instance, it is still unclear how the stability of two-component BECs influences the properties of polarons, and how the two Bogoliubov excitations contribute to the characteristics of polarons.

In this paper, we aim to explore the properties of Bose polarons formed by coupling impurities with two Bogoliubov excitations in two-component BECs. Specifically, we will investigate the ground state energy, effective mass, and average number of virtual phonons. To begin, we derive an effective Fr\"{o}hlich Hamiltonian using the Bogoliubov approximation and the generalized Bogoliubov transformation. This derivation allows us to define two types of effective interactions: impurity-density coupling and impurity-spin coupling. Next, we employ the Lee-Low-Pines variational theory to derive expressions for the relevant properties of Bose polarons. Our analysis reveals an interesting finding: as the intercomponent interaction approaches the unstable regime's edge, the impurity-spin coupling becomes significantly stronger. As a result, the corresponding ground state energy decreases to zero. Additionally, a large number of phonons are induced, which impedes the motion of the impurities and leads to a significant increase in the effective mass of the Bose polaron. On the other hand, we observe that the contribution of impurity-density coupling to the properties of polarons closely resembles what is observed in single-component BECs. By investigating these properties, we hope to gain a deeper understanding of polarons and their behavior in two-component BECs.

The structure of the remainder of this paper is as follows. In Sec. \ref{s2}, we propose a Hamiltonian to characterize the Bose polaron formed by the interaction of an impurity and a two-component BEC. The derivation of the polaron properties from the Lee-Low-Pines variational theory is presented in Sec. \ref{s3}. In Sec. \ref{r1}, we provide a detailed calculation for the case of a homonuclear BEC, while in Sec. \ref{r2}, we present the calculation for the case of a heteronuclear BEC. Finally, we conclude with a summary in Sec. \ref{c}.

\section{Effective Hamiltonian}\label{s2}
In this section, we present a model for a slow-moving impurity immersed in a binary BEC. The binary BEC comprises three-dimensional, homogeneous two-component Bose gases. By employing the Bogoliubov approximation, we derive a Fr\"{o}hlich-like Hamiltonian~\cite{Frohlich1954}, disregarding higher-order correlations of Bogoliubov phonons. This approximation reveals a weak coupling between the impurity and Bogoliubov excitations. The system we investigate in this paper is akin to those discussed in previous studies, such as~\cite{Charalambous2020}, except for the absence of the Rabi coupling between the two components. Moreover, The binary BEC is not restricted to homonuclear bosons but encompasses heteronuclear bosons as well. While the experimental implementation necessitates two distinct schemes, the underlying theoretical model remains the same~\cite{Pita2016}.

We consider the Hamiltonian of the entire system, which consists of three parts:
\begin{equation}
H=H_{I}+H_{B}+H_{IB}.\label{H}
\end{equation}
Here, $H_I=p_I^2/2m_I$ represents the kinetic energy of the impurity, where
 $m_I$ is the mass of the impurity. $H_B$ characterizes the two-component BECs, given by
\begin{equation}
  H_B=H_1+H_2+H_{12}.\label{HB}
\end{equation}
The terms $H_j$ in $H_B$ describe the binary BEC with intracomponent interaction,
\begin{equation}
  H_j=\sum_{\bm{k}}\epsilon_{j}a^\dagger_{j,\bm{k}}a_{j,\bm{k}}+\frac{1}{2\Omega}\sum_{\bm{pqk}}V_{jj}a^\dagger_{j,\bm{p}+\bm{k}}a^\dagger_{j,\bm{\bm{q}-\bm{k}}}a_{j,\bm{q}}a_{j,\bm{p}},\label{Hj}
\end{equation}
Here, $\Omega$ is the volume of the condensates. $\epsilon_{j}=\hbar^2 k^2/(2m_j)$ and $m_j$ are the kinetic energy and mass for each component respectively. Here, $j=1, 2$.  The term of $H_{12}$ represents the intercomponent interaction for the two-component BEC,
\begin{equation}
  H_{12}=\frac{1}{\Omega}\sum_{\bm{pqk}}V_{12}a^\dagger_{1,\bm{p}}a_{1,\bm{p}+\bm{k}}a^\dagger_{2,\bm{q}}a_{2,\bm{q}-\bm{k}}.
\end{equation}
In addition, $V_{jj}$ represents intracomponent interactions. $V_{12}$ is the intercomponent interaction. The term $H_{IB}$ represents the Fr\"{o}hlich term that couples the impurity with the bosons,
\begin{equation}
  H_{IB}=\frac{1}{\sqrt{\Omega}}\sum_{j,\bm{pk}}e^{{\rm i}\bm{k}\cdot \bm{r}}V_{jI}(\bm{k})a^\dagger_{j,\bm{p}-\bm{k}}a_{j,\bm{p}}.
\end{equation}
 $\bm{r}$ is the position vector of the impurity. $e^{{\rm i}\bm{k}\cdot\bm{r}}=\rho(\bm{k})=\int d\bm{r}'e^{{\rm i}\bm{k}\cdot\bm{r}}\delta(\bm{r}-\bm{r}')$ represents the impurity density in momentum space. The multiple interactions $V_{jj}, V_{12}$, and $V_{jI}$ depend on the scattering length $a_{jj}, a_{12}$, and $a_{jI}$, respectively. Specifically, $V_{jj}=4\pi\hbar^2 a_{jj}/m_j, V_{12}=2\pi\hbar^2 a_{12}/m_{12}$, and $V_{jI}=4\pi\hbar^2 a_{jI}/m_{jI}$, where $m_{12}={m_1m_2}/({m_1+m_2})$ and $ m_{jI}={m_jm_I}/({m_j+m_I})$.

\begin{figure}[!htpb]
\centering
\includegraphics[width=18cm]{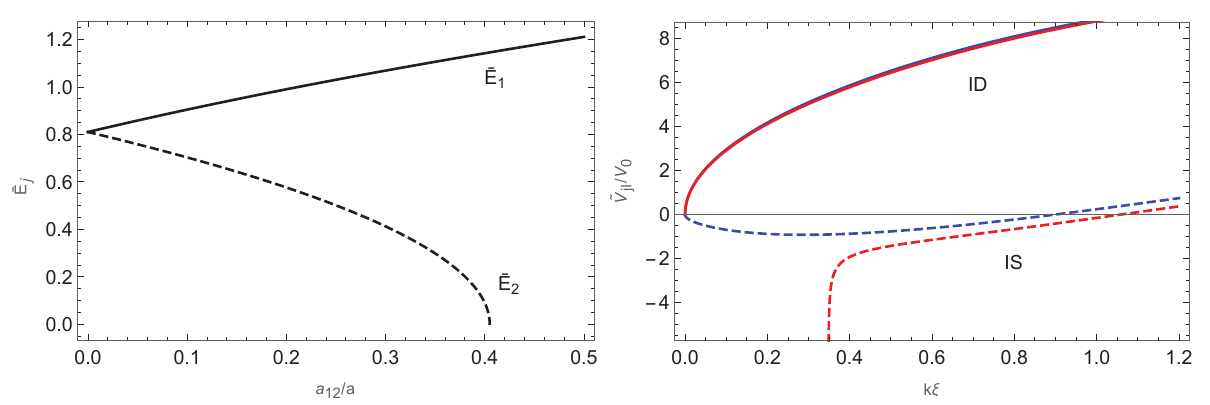}
\caption{(color online) (a) The excitation spectrum as a function of $a_{12}/a$, where $\bar{E}_{j}=\bar{k}\sqrt{\bar{k}^2[1+(-1)^{j-1}2a_{12}/a]}$ is obtained from Eq.(\ref{ce}). Here $\bar{k}=k\xi=0.9$ with the healing length $\xi=1/\sqrt{8\pi a n}$. Other parameters are set as $m_1=m_2$, $a=a_{11}=a_{22}$, and $n=n_{1,0}=n_{2,0}$. (b) illustrates ID and IS couplings corresponding to $\tilde{V}_{1I}$ and $\tilde{V}_{2I}$ respectively. The parameter $V_0=2\pi\hbar^2a_{1I}/m_I$ is used. In this case, the mass parameters are chosen as $m_2=2m_1$, and $m_I=10m_1$. The blue curves correspond to the scattering length of $a_{12}=0.9a$, while the red curves correspond to $a_{12}=a$. }\label{v12}
\end{figure}

Using the Bogoliubov approximation $a_{j,0}\rightarrow \sqrt{N_{j,0}}$, the quartic terms in Eq.(\ref{HB}) can be transformed into quadratic terms. The intracomponent interaction term $H_j$ becomes
\begin{equation}
    H_j=\sum_{\bm{{k}}}\left(\epsilon_{j}+N_{j,0}V_{jj}+N_{j,0}V_{12}\right)a^\dagger_{j,\bm{k}}a_{j,\bm{k}}+\sum_{\bm{{k}}}\frac{n_{j,0}}{2}V_{jj}\left(a_{j,-\bm{k}}a_{j,\bm{k}}+a^\dagger_{j,\bm{k}}a^\dagger_{j,-\bm{k}}\right),
  \end{equation}
where $n_{j,0}=N_{j,0}/\Omega$ represents the number density of condensate particles for the j component. The intercomponent interaction term $H_{12}$ becomes,
  \begin{equation}
    H_{12}= \sum_{\bm{{k}}}\sqrt{n_{1,0}n_{2,0}}V_{12}\left(a_{1,\bm{k}}+a^\dagger_{1,-\bm{k}}\right)\left(a_{2,-\bm{k}}+a^\dagger_{2,\bm{k}}\right).
  \end{equation}
 Finally, the term $H_{IB}$ is reduced to a linear form,
  \begin{equation}
     H_{IB}=\sum_jV_{jI}N_{j,0}+\sum_{j,\bm{{k}}}\sqrt{n_{j,0}}V_{jI}e^{i\bm{k}\cdot\bm{r}}(a^\dagger_{j,-\bm{k}}+a_{j,\bm{k}}).\label{HIB}
  \end{equation}
To obtain the diagonalized effective Hamiltonian, we need to apply a general Bogoliubov transformation. In this model, without the Rabi coupling, the rotation step in the general Bogoliubov transformation is not required to eliminate the Rabi coupling term, as shown in previous works~\cite{Tommasini2003, Charalambous2020}.  By introducing the generalized Bogoliubov transformation~\cite{Larsen1963, Sun2010, Ota2020}, given by
\begin{subequations}
\begin{align}
  a_{1,\bm{k}}&=U^{+}_{11,k}b_{1,\bm{k}}+U^{-}_{11,{k}}b^\dagger_{1,-\bm{k}}-(U^{+}_{12,{k}}b_{2,\bm{k}}+U^{-}_{12,{k}}b^{\dagger}_{2,-\bm{k}}),\\
  a_{2,\bm{k}}&=U^{+}_{22,k}b_{2,\bm{k}}+U^{-}_{22,{k}}b^\dagger_{2,-\bm{k}}+(U^{+}_{21,{k}}b_{1,\bm{k}}+U^{-}_{21,{k}}b^{\dagger}_{1,-\bm{k}}).
\end{align}\label{bt}
\end{subequations}
The Bogoliubov transform establishes the relationship between the particle number operator $a_{j,\bm{k}}$ and the quasiparticle operator $b_{j,\bm{k}}$. The inverse Bogoliubov transformation can also be obtained by Eq.(\ref{bt}). To diagonalize the Hamiltonian, the coefficients $U^{\pm}_{ij,k}$ are given by
\begin{subequations}
\begin{align}
U^{\pm}_{jj,k}&= \frac{1}{2}\left(\sqrt{\frac{\epsilon_{j}}{E_{j}}}\pm\sqrt{\frac{E_{j}}{\epsilon_{j}}}\right)\cos\gamma_k \\U^{\pm}_{jj',k}&=\frac{1}{2}\left(\sqrt{\frac{\epsilon_{j}}{E_{j'}}}\pm\sqrt{\frac{E_{j'}}{\epsilon_{j}}}\right)\sin\gamma_k
\end{align}
\end{subequations}
where any $j\ne j'$. The $\gamma_k$ and the collective excitations $E_{j}$ are given by
\begin{equation}
\cos\gamma_k=\sqrt{\frac{1}{2}(1+x)},  \ \sin\gamma_k= \sqrt{\frac{1}{2}(1-x)},
\end{equation}
\begin{equation}
x=\frac{\varepsilon^2_{1}-\varepsilon^2_{2}}{\sqrt{(\varepsilon^2_{1}-\varepsilon^2_{2})^2+16V^2_{12}\epsilon_{1,k}\epsilon_{2,k}n_{1,0}n_{2,0}}}.
\end{equation}
\begin{equation}
   E_{j}^2=\frac{1}{2}(\varepsilon_{1}^2+\varepsilon_{2}^2)+(-1)^{j-1}\frac{1}{2}\sqrt{(\varepsilon_{1}^2-\varepsilon_{2}^2)^2+16V^2_{12}\epsilon_{1}\epsilon_{2}n_{1,0}n_{2,0}}\label{ce},
\end{equation}
where
\begin{align}
  \varepsilon^2_{j}=\epsilon^2_{j}+2V_{jj}n_{j,0}\epsilon_{j}.
\end{align}
Both collective excitations $E_{1}$ and $E_{2}$ correspond to the gapless modes associated with Bogoliubov density and spin fluctuations, respectively~\cite{Ota2020}. Fig.\ref{v12}(a) illustrates the distinct behaviors of two kinds of excitations in terms of intercomponent interactions. The diagonalized Hamiltonian $H_d$ is expressed as
\begin{equation}
  H_d=\frac{p_I^2}{2m_I}+\sum_{j,\bm{k}}E_{j}b^\dagger_{j,\bm{k}}b_{j,\bm{k}}+\sum_jV_{jI}N_{j,0}+\sum_{j,\bm{k}}\sqrt{\frac{n_{j,0}}{\Omega}}\tilde{V}_{jI}e^{i\bm{k}\cdot\bm{r}}(b_{j,\bm{k}}+b_{j,-\bm{k}}^\dagger),\label{Hd}
\end{equation}
where $\tilde{V}_{jI}$ is defined as
\begin{equation}
  \tilde{V}_{jI}=V_{jI}\sqrt{\frac{\epsilon_{j}}{E_{j}}}\cos\gamma_k+(-1)^{j-1}V_{j'I}\sqrt{\frac{\epsilon_{j'}}{E_{j}}}\sin\gamma_k\label{ev},
\end{equation}
with any $j'\ne j$ in (\ref{ev}). By employing the definition of $\tilde{V}_{jI}$ in Eq.(\ref{ev}), the model of an impurity immersed in the binary BEC can be mapped onto the conventional Fr\"{o}hlich model. The effective interactions $\tilde{V}_{jI}$ correspond to linear transformations of the original interactions, featuring a momentum-dependent structure. Specifically, $\tilde{V}_{1I}$ solely depends on $E_{1}$ and is referred to as the impurity-density (ID) coupling, while $\tilde{V}_{2I}$ depends solely on $E_{2}$ and is termed the impurity-spin(IS) coupling.

We demonstrate the distinct behaviors of the ID and IS coupling in Fig. \ref{v12}(b). The ID coupling exhibits only repulsive interactions, which monotonically increase with increasing $|\bm{k}|$. On the other hand, the behavior of the IS coupling is not simply monotonically growing. For
$a_{12}/a=0.9$, the IS coupling corresponds to the blue dashed line in Figure \ref{v12}(b), and we observe that the IS coupling decreases at first and then increases. Moreover, the IS coupling induces an attractive interaction for small $|\bm{k}|$. It is important to note that the IS coupling behaves diversely, which means that binary BECs exhibit instability. The instability comes from the violation that $E_2$ must be real. The condition for instability is given by
\begin{equation}
 (\epsilon_{1}+2V_{11}n_{1,0})(\epsilon_{2}+2V_{22}n_{2,0}) \le 4V_{12}^2n_{1,0}n_{2,0}\label{sc}.
\end{equation}
For small $|\bm{k}|$ and the same density, Eq.(\ref{sc}) becomes
\begin{equation}
  V_{11}V_{22}\le V_{12}^2.
\end{equation}
This is the condition of phase separation \cite{Pitaevskii2016}. It implies that IS couplings make no sense in the immiscible region. We show the complete behavior of effective interaction in Appendix \ref{A1}. It can be observed that the boundary line between the stable and unstable regions of IS couplings is described by Eq.(\ref{sc}).

\begin{figure}[!htpb]
\centering
\includegraphics[width=9cm]{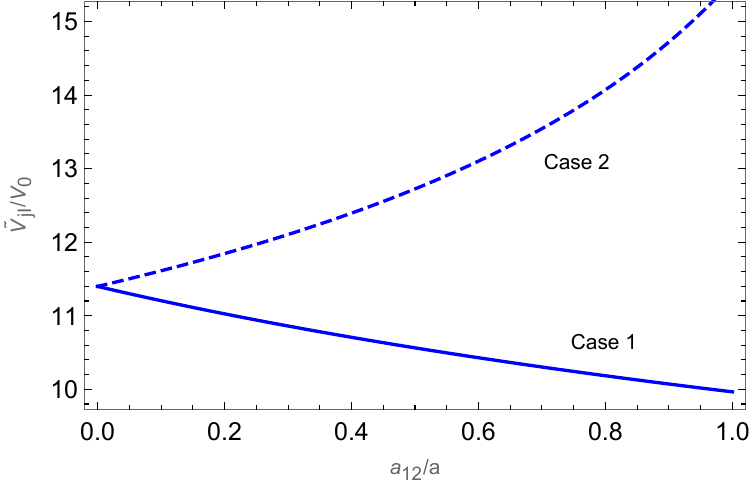}
\caption{(color online) Effective interactions in case 1 and case 2 respectively, with $\bar{k}=0.9$ and $m_I=10m$.}\label{vc12}
\end{figure}

To facilitate the understanding the properties of polarons formed by the effective interactions of impurities with Bogoliubov phonons, we consider two scenarios that are similar to those presented in~\cite{Charalambous2020}. In both scenarios, we assume that the mass and scattering lengths are equal, with $m_1=m_2$, $a=a_{11}=a_{22}$. The difference between two scenarios lies in the boson-impurity scattering lengths: in Case 1, the boson-impurity scattering lengths are the same, $a_{1I}=a_{2I}$, while in Case 2, they are opposite,
 $a_{1I}=-a_{2I}$. Consequently, in Case 1, the IS interaction vanishes, while ID interactions remain. The last term of Eq.(\ref{Hd}) becomes
\begin{equation}
  {\rm Case \ 1}: \quad  \sum_{\bm{k}}\sqrt{\frac{n_{j,0}}{\Omega}}\tilde{V}_{1I}e^{i\bm{k}\cdot\bm{r}}(b_{1,\bm{k}}+b_{1,-\bm{k}}^\dagger), \ {\rm where} \quad \tilde{V}_{1I}={V}_{BI}\sqrt{\frac{2\epsilon_{k}}{E_{1}}}.\label{c1}
\end{equation}
Here $\epsilon_k=\epsilon_{1}=\epsilon_{2}$ and ${V}_{BI}={V}_{1I}={V}_{2I}$. In Case 2, on the other hand, the ID interaction vanishes, while IS couplings remain. The last term of Eq.(\ref{Hd}) becomes
\begin{equation}
  {\rm Case \ 2}: \quad  \sum_{\bm{k}}\sqrt{\frac{n_{j,0}}{\Omega}}\tilde{V}_{2I}e^{i\bm{k}\cdot\bm{r}}(b_{2,\bm{k}}+b_{2,-\bm{k}}^\dagger), \ {\rm where} \quad \tilde{V}_{2I}={V}_{BI}\sqrt{\frac{2\epsilon_{k}}{E_{2}}}.\label{c2}
\end{equation}
Here ${V}_{BI}={V}_{2I}=-{V}_{1I}$. Therefore, in conjunction with Fig.~\ref{v12}(a), we find that the ID coupling decreases with the intercomponent interaction, whereas the IS coupling increases. The effective interaction in both cases is shown in Fig.\ref{vc12}.

 We need to discuss the validity of the Bogoliubov approximation, which assumes that the quantum depletion, denoted as $\sum_{\bm{k}}\langle a_{j,\bm{k}}^\dagger a_{j,\bm{k}}\rangle$, is negligible. By employing the generalized Bogoliubov transformation in Eq.(\ref{bt}), we can obtain both the diagonal term $\langle b^\dagger_{j,\bm{k}}b_{j,\bm{k}}\rangle$  and the off-diagonal terms such as $\langle b_{j,\bm{k}}b_{j,\bm{k}}\rangle$  that contribute to the quantum depletion. The diagonal term arises from the interaction between the Bose atoms, while the off-diagonal terms are influenced by the presence of impurities, primarily depending on the interaction between the impurities and bosons, as well as the ratio of impurity number to boson number. In the case of weak intercomponent interactions and a small number of impurity atoms, the quantum depletion is minimal. Therefore, the total quantum depletion satisfies the following condition:
 \begin{equation}
   \sum_{j,\bm{k}}\langle a_{j,\bm{k}}^\dagger a_{j,\bm{k}}\rangle/N_{j}\ll 1.
 \end{equation}
Here, $N_j$ represents the total number of particles in each component. This condition confirms the validity of the Bogoliubov approximation within this model. The diagonal term of $\tilde{n}_j$ has been studied in~\cite{Tommasini2003,Eckardt2004}. The analysis presented in~\cite{Boudjemaa2020} evaluated the impact of impurities on the quantum depletion of the Bose-Bose mixture and found that the depletion is very small.

\section{The Variational Approach}\label{s3}
In this section, we extend the analysis of the Lee-Low-Pines variational theory from the single-component BEC case~\cite{Huang2009} to the case of binary BECs. We investigate various properties of Bose polarons, including the average number of virtual phonons, ground state energy, and effective mass.

To eliminate the position operator $\bm{r}$ of the impurity in Eq.(\ref{Hd}), we employ the Lee-Low-Pines transformation, $\tilde{H}=S^{-1}H_dS$, where
\begin{equation}
  S=\exp{\left(-i\sum_{j,\bm{k}}b_{j,\bm{k}}^\dagger b_{j,\bm{k}}\bm{k}\cdot \bm{r}\right).}\label{s1}
\end{equation}
Following several calculations, we obtain the effective Hamiltonian as
\begin{equation}
\begin{aligned}
  \tilde{H}&=\sum_jV_{jI}N_{j,0}+\frac{1}{2m_{I}}(\bm{p}-\sum_{j,\bm{k}}\hbar\bm{k}b_{j,\bm{k}}^\dagger b_{j,\bm{k}})^2+\sum_{j,\bm{{k}}}\left[E_{j}b^\dagger_{j,\bm{k}}b_{j,\bm{k}}+\sqrt{\frac{N_{j,0}}{\Omega}}\tilde{V}_{jI}(b_{j,\bm{k}}^\dagger+b_{j,\bm{k}})\right].\label{eh}
\end{aligned}
  \end{equation}
Here, $\bm{p}$ can be considered as a c number after the transformation Eq.(\ref{s1}), i.e., $\bm{p}$ represents the total momentum of the system. We adopt the ansatz wave function,
\begin{equation}
  |\psi\rangle=U|\psi_0\rangle=e^{\sum_{j,\bm{k}}\left(b_{j,\bm{k}}^\dagger \beta_j(k)-b_{j,\bm{k}} \beta^*_j(k)\right)}|\psi_0\rangle,
  \label{An}
\end{equation}
where $|\psi_0\rangle$ represents the common phonon vacuum state for the binary Bose system. The unitary transformation $e^{\sum_{j,\bm{k}}\left(b_{j,\bm{k}}^\dagger \beta_j(k)-b_{j,\bm{k}} \beta^*_j(k)\right)}$ corresponds to a displacement operator for $b_{j,\bm{k}}$ and $b_{j,\bm{k}}^\dagger$. The variational parameter $\beta_{j}(k)$ is utilized to minimize the variational energy. The variational ansatz given in Eq.(\ref{An}) originates from the pioneering work of Lee, Low, and Pines~\cite{Lee1953}, representing an approximation of coherent states. However, it is exact for impurity with infinite mass~\cite{Shchadilova2016}. The variational wave function in Eq.(\ref{An}) incorporates the emission of any number of virtual phonons, enhancing the accuracy of the Lee-Low-Pines theory in comparison to the perturbation theory. It is important to note that the emission of virtual phonons is statistically independent, implying the disregard of phonon correlations.

We note the annihilation operator $b_{j,\bm{k}}|\psi_0\rangle=0$, and the normalization condition $\langle\psi_0|\psi_0\rangle=1$ holds. Consequently, the energy functional $E_p=\langle \psi_0|U^{-1}\tilde{H}U|\psi_0\rangle$ is expressed as
\begin{equation}
\begin{aligned}
  E_p&=\sum_jV_{jI}N_{j,0}+\left(1-\eta\right)^2\frac{\bm{p}^2}{2m_I}+\sum_{j,\bm{k}}\sqrt{n_{j,0}}\tilde{V}_{jI}(\beta_{j}({k})+\beta^*_j(k))+\sum_{j,\bm{k}}|\beta_{j}(k)|^2\left(E_{j}+\frac{\hbar^2k^2}{2m_I}\right).\end{aligned}\label{Ef}
\end{equation}
Here, the parameter $\eta$  is defined as
 \begin{equation}
 \eta \bm{p}=\sum_{j,\bm{k}}|\beta_{j}(k)|^2\hbar\bm{k}.\label{emp}
 \end{equation}
At the energy functional minimum, $\delta E_p/\delta\beta _j{(k)}=\delta E_p/\delta\beta^{*} _j{(k)}=0$, and the expression for $\beta_j(k)$ is given by
\begin{equation}
  \beta_j(k)=-\frac{\sqrt{n_{j,0}}\tilde{V}_{jI}}{E_{j}-(1-\eta)\frac{\bm{p}}{m_I}\cdot\hbar\bm{k}+\frac{\hbar^2k^2}{2m_I}}.\label{vp}
\end{equation}
Since the impurities in this model move slowly, $\bm{p}$ is a small quantity, allowing us to assume that:
\begin{equation}
(1-\eta)\frac{\bm{p}}{m_I}\cdot\hbar \bm{k}\ll E_{j}+\frac{\hbar^2k^2}{2m_I}.\label{ap}
\end{equation}
By using Eq.(\ref{ap}) and Eq.(\ref{vp}), we can obtain from Eq.(\ref{emp})
\begin{equation}
 W= \frac{\eta}{1-\eta}=\sum_{j,\bm{k}}
\frac{n_{j,0}\tilde{V}_{jI}^2}{\left(E_{j}+\frac{\hbar^2k^2}{2m_I}\right)^3}\frac{2\hbar^2 }{m_I}k^2\cos^2\theta\label{em},
\end{equation}
where $\cos\theta$ arises from the dot product of $\bm{p}$ and $\bm{k}$. By inserting Eq.(\ref{vp}) into Eq.(\ref{Ef}) and considering the approximation in Eq.(\ref{ap}), we obtain
  \begin{equation}
  E_p(\bm{p})=E_p(0)+\frac{p^2}{2m^*}+\mathcal{O}(p^4).\label{fe}
\end{equation}
The first term $E_p(0)$ represents the ground state energy
 with $\bm{p}=0$,
\begin{equation}
  E_p(0)=\sum_jV_{jI}N_{j,0}-\sum_{j,\bm{k}}\frac{n_{j,0}\tilde{V}_{jI}^2}{E_{j}+\frac{\hbar^2k^2}{2m_I}}\label{E}.
\end{equation}
The second term on the right-hand side of Eq.(\ref{fe}) defines an effective mass $m^*$, with $m^*=m_I/(1-\eta)$. We introduce the ratio of the effective mass to the mass of the impurity, $m^*/m_I=1+W$, where $W$ is the left-hand side of Eq.(\ref{em}). By evaluating $W$, we can determine the effective mass of the polaron. Once $W$ and $E_p(0)$ are obtained, the energy of the polaron given in Eq.(\ref{fe}) can be determined. In the following, we will use $E_p$ instead of $E_p(0)$ for brevity.

A polaron can be visualized as an impurity dressed in phonon-like excitations. Therefore, the average number of virtual phonons also plays a crucial role in characterizing polarons. It is important to highlight that the process of phonon emission by impurities is not a real process and does not adhere to the conservation of energy. When $\bm{p}=0$, the expression for the average number of virtual phonons can be obtained as follows:
\begin{align}
  N_{\rm ph}=\sum_{j,k}|\beta_{j}(k)|^2=\sum_{j,k}\frac{n_{j,0}\tilde{V}_{jI}^2}{\left(E_{j}+\frac{\hbar^2k^2}{2m_I}\right)^2}\label{N}.
\end{align}
The results of the relevant properties are presented in Sec.\ref{r}.

\section{Results}\label{r}
In this section, we present a comprehensive analysis of the ground state energy, effective mass, and average number of virtual phonons. However, it is important to note that the calculations involve complex expressions, and therefore, we have made certain simplifying assumptions to facilitate the analysis.

For a two-component homonuclear BEC, where the Bose atoms have the same mass, we investigate two specific cases mentioned at Sec. \ref{s2} to compute the relevant polaronic properties. It should be noted that the integrals of the ground state energy exhibit divergence due to the artificial selection of contact interactions. However, this divergence can be eliminated by considering higher-order contributions of the scattering length.

In the case of a two-component heteronuclear BEC, where the masses of the Bose atoms are not equal, we consider the weak intercomponent interaction to derive the formulas influenced by the mass imbalance of Bose atoms. The obtained integral results cannot be directly deduced from the formula of a single-component BEC. Therefore, the results obtained are non-trivial.
\subsection{Mass Balance}\label{r1}

We consider a binary BEC where two components have equal mass, $m=m_1=m_2$, and each component occupies the same density, $n=n_{1,0}=n_{2,0}$. By converting the summation over
$\bm{k}$ in Eq.(\ref{E}) to an integral, $\frac{1}{\Omega}\sum_{\bm{k}}\rightarrow \frac{1}{(2\pi)^3}\int d\bm{k}$, we obtain the expression for the single polaron energy as follows:
\begin{equation}
  E_p(0)=V_{1I}n+V_{2I}n+E_p^++E_p^-,\label{Em}
\end{equation}
where
\begin{subequations}
\begin{align}
E_p^{\pm}&=-\frac{E_{0}}{\sqrt{3\pi^3}}\frac{(k_n a_{1I})^2}{(k_na_{11})^{1/2}}(1+\bar{m})^2I^{\pm}_E,\\
 I^{+}_E&=\int_0^\infty {\rm d}\bar{k}\frac{\left(\sqrt{1+\bar{a}}+\frac{a_{2I}}{a_{1I}}\sqrt{1-\bar{a}}\right)^2\bar{k}^2}{\bar{k}^2+a_{+}+\bar{m}\bar{k}\sqrt{\bar{k}^2+a_{+}}},\\
 I^{-}_E
 &=\int_0^\infty {\rm d}\bar{k}\frac{\left(\sqrt{1-\bar{a}}-\frac{a_{2I}}{a_{1I}}\sqrt{1+\bar{a}}\right)^2\bar{k}^2}{\bar{k}^2+a_{-}+\bar{m}\bar{k}\sqrt{\bar{k}^2+a_{-}}},\label{I}
\end{align}
\end{subequations}
where $\bar{m}=m/m_I$, $E_{0}=2\pi\hbar^2a_{11} n/m$, $\bar{k}=k\xi_{11}$, $\xi_{11}=1/\sqrt{8\pi a_{11}n }$, and $k_n^3=6\pi^2 n$. The dimensionless quantities $\bar{a}$ and $a_{\pm}$ contain all kinds of scattering lengths of the binary BECs, given by
\begin{align}
  \bar{a}=\frac{1}{\sqrt{\Delta}}(a_{11}-a_{22}), \quad a_{\pm}=1+\frac{1}{a_{11}}\left(a_{22}\pm \sqrt{\Delta}\right),\label{dl}
\end{align}
where $\Delta=(a_{11}-a_{22})^2+4a_{12}^2$. The condition $a_->0$, which implies $a_{11}a_{22}-a_{12}^2>0$, ensures the stability of binary BECs, while $a_-<0$ indicates phase separation. Next, we consider two typical cases to illustrate the contribution of ID and IS couplings to the ground energy. These cases correspond to obtaining an effective interaction in Eq.(\ref{c1}) and (\ref{c2}).

In Case 1, the contribution of IS couplings to the energy $E_p^-$ vanishes, with $\bar{a}=0$, $I^-_E=0$, and $a_{\pm}=2(1\pm a_{12}/a)$.
The energy of the polaron becomes $E_p(0)=V_{1I}n+V_{2I}n+E_p^+$, where
\begin{align}
  E_p^+=-\frac{E_0}{\sqrt{3\pi^3}}\frac{(k_n a_{1I})^2}{(k_na)^{1/2}}(1+\bar{m})^2I^{+}_{E},\label{Ep}
  \end{align}
Noting that $I_E^+$ diverges like $\sum_{\bm{k}}k^{-2}$, an ultraviolet (UV) cutoff is needed in the summation. However, the second-order scattering contribution of $V_{1I}n_{1,0}$ also diverges in the same way. To eliminate the divergence in the integral $I^{+}_E$, we absorb the second term on the right-hand side of
\begin{equation}
 \frac{2\pi \hbar^2a_{1I}n}{m_{1I}}= \left(V_{1I}n-\frac{2m_{1I}V_{1I}^2n}{\hbar^2}\sum_{\bm{k}}\frac{1}{\bm{k}^2}\right)\label{ci}
\end{equation}
Therefore, the divergent $I_{E}^+$ is replaced by the convergent $I^{+}_{EC}$, given by
\begin{align}
 I^{+}_{EC}=4\int_0^\infty {\rm d}\bar{k}\left(\frac{1}{1+\bar{m}}-\frac{\bar{k}^2}{\bar{k}^2+a_++\bar{m}\bar{k}\sqrt{\bar{k}^2+a_+}}\right).
\end{align}
The integral result can be expressed as
\begin{equation}
 I^{+}_{EC}=4\sqrt{2\left(1+ \frac{a_{12}}{a}\right)}\left[\frac{ \arccos\bar{m}}{(1-\bar{m}^2)^{3/2}}-\frac{\bar{m}}{1-\bar{m}^2}\right].\label{EI}
\end{equation}
The expression of $E_p^+$ is given by
\begin{equation}
  E_p^{+}/E_0=\frac{8}{\sqrt{6\pi^3}}\frac{(k_n a_{1I})^2}{(k_na)^{1/2}}(1+\bar{m})^2\sqrt{\left(1+ \frac{a_{12}}{a}\right)}\left[\frac{ \arccos\bar{m}}{(1-\bar{m}^2)^{3/2}}-\frac{\bar{m}}{1-\bar{m}^2}\right].\label{E+}
\end{equation}
The integral result can be reduced to the case of the single-component BEC. when $a_+=2$ (i.e., $a_{12}=0$), which is consistent with the result in Ref.~\cite{Novikov2009}. It is worth noting that the second term of the polaron energy has a structure of the Lee-Huang-Yang term when all scattering lengths are equal in the system consisting of bosons and an impurity~\cite{Novikov2009, Fetter1971}.

In Case 2, the contribution of ID couplings to the energy $E_p^+$ vanishes. Here, $\bar{a}=0$, $a_{\pm}=2(1\pm a_{12}/a)$, and $I^+_E=0$. The energy of the polaron in Eq.(\ref{Em}) becomes $E_p(0)=V_{1I}n+V_{2I}n+E_p^-$.
Following the same process as in Case 1, we obtain similar integral results, except that $a_{+}$
  in Eq.(\ref{EI}) needs to be replaced by $a_{-}$, i.e.,
 \begin{equation}
 I^{-}_{EC}=\sqrt{2\left(1- \frac{a_{12}}{a}\right)}\left[\frac{ \arccos\bar{m}}{(1-\bar{m}^2)^{3/2}}-\frac{\bar{m}}{1-\bar{m}^2}\right].\label{EIS}
\end{equation}
The result of $E_p^-$ is given by
\begin{equation}
  E_p^{-}/E_0=\frac{8}{\sqrt{6\pi^3}}\frac{(k_n a_{1I})^2}{(k_na)^{1/2}}(1+\bar{m})^2\sqrt{\left(1- \frac{a_{12}}{a}\right)}\left[\frac{ \arccos\bar{m}}{(1-\bar{m}^2)^{3/2}}-\frac{\bar{m}}{1-\bar{m}^2}\right].\label{E-}
\end{equation}

\begin{figure}[ht]
\centering
\includegraphics[width=16cm]{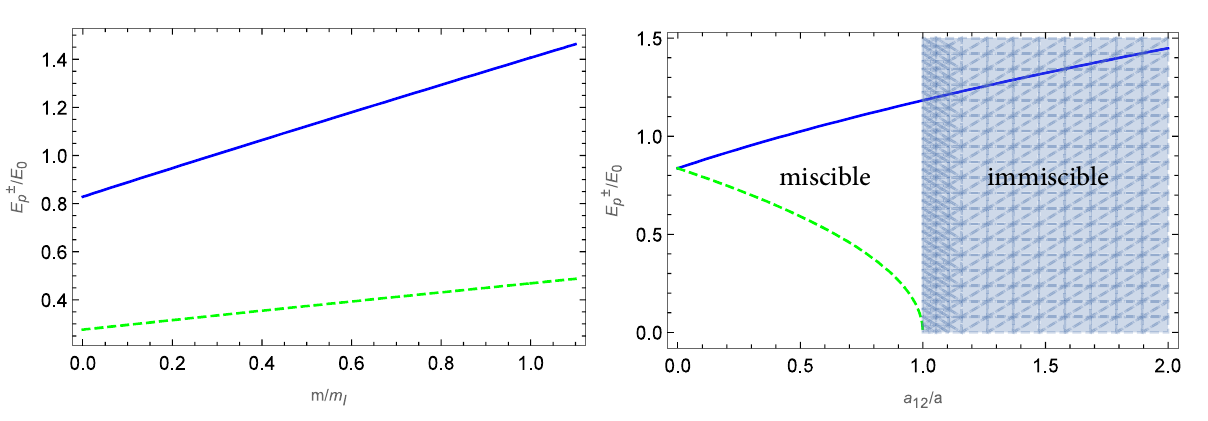}
\caption{Polaron energy as a function of $m/m_I$ (a) with $a_{12}/a=4/5$, and as a function of $a_{12}/a$ (b) with $m/m_I=1/2$. The solid line represents $E_p^+$, i.e., Eq.(\ref{E+}). The dashed line represents $E_p^-$, i.e, Eq.(\ref{E-}). The parameters are set as $k_na=0.9$ and $k_na_{1I}=0.8$.}\label{Ema}
\end{figure}

We present the energy plots of $E_p^+$ (solid line) and $E_p^-$ (dashed line) as functions of $m/m_I$ in Fig.\ref{Ema}(a), and as a function of $a_{12}/a$ in Fig.\ref{Ema}(b). In Fig.\ref{Ema}(a), Both $E_p^+$ and $E_p^-$ exhibit similar trends, differing only in the coefficients that depend on $m/m_I$. Fig.\ref{Ema}(b) demonstrates the inverse relationship between the behaviors of $E_p^+$ and $E_p^-$. Specifically, as $a_{12}/a$ increases, the energy induced by the ID coupling increases while the energy induced by the IS coupling decreases. In Fig.\ref{Ema}(b), the region where $a_{12}/a<1$ represents the immiscible region, while $a_{12}/a>1$ corresponds to the miscible region. It is important to note that at the critical value $a_{12}/a=1$, $E^-_p$ becomes zero. This implies that in the immiscible region, there is no contribution of IS couplings to the polaron energy. This is due to the inability of the impurity to simultaneously interact with both components in the immiscible region. Therefore, in the subsequent analysis of the effective mass and the average number of virtual phonons, we only consider the miscible region.

We convert the summations over $\bm{k}$ in (\ref{em}) and Eq.(\ref{N}) to integrals and observe that they converge. To calculate the effective mass of the polaron and the average number of phonons, we consider two cases, which is the same process used to calculate the energy. In Case 1, we only analyze the contribution of $W^+$ ($N^+_{\rm ph}$) from ID couplings, while in Case 2, the contribution of $W^-$ ($N^-_{\rm ph}$) remains. We present the contribution of ID and IS couplings together in the expressions, but it should be noted that they are obtained respectively in Case 1 and Case 2. The effective mass of the polaron and the average number of phonons in equal-mass binary BECs are as follows:
\begin{align}
  W^{\pm}&=\frac{8}{3\sqrt{3\pi^3}}\frac{(k_B a_{1I})^2}{(k_Ba)^{1/2}}\bar{m}(1+\bar{m})^2I_W^{\pm},\\
   N^{\pm}_{\rm ph}&=\frac{2}{\sqrt{3\pi^3}}\frac{(k_B a_{1I})^2}{(k_Ba)^{1/2}}(1+\bar{m})^2I^{\pm}_N,\label{N}
\end{align}
where
\begin{align}
  I_W^{\pm}&=\int_0^\infty {\rm d}\bar{k}\frac{\bar{k}^2}{\sqrt{\bar{k}^2+a_{\pm}}}\frac{1}{\left(\sqrt{\bar{k}^2+a_{\pm}}+\bar{m}\bar{k}\right)^3},\label{IW}
  \\
  I^{\pm}_N&=\int_0^\infty{\rm d}\bar{k}\frac{\bar{k}}{\sqrt{\bar{k}^2+a_{\pm}}}\frac{1}{\left(\sqrt{\bar{k}^2+a_{\pm}}+\bar{m}\bar{k}\right)^2}.
\end{align}
The analytical results of the integrals are given by,
\begin{align}
  I_W^{\pm}&=\frac{1}{2\sqrt{2\left(1\pm {a_{12}}/{a}\right)}}\left[\frac{1+2\bar{m}^2}{(1-\bar{m}^2)^{5/2}}\arccos \bar{m}-\frac{3\bar{m}}{(1-\bar{m}^2)^2}\right].\label{WI}\\
  I^{\pm}_N&=\frac{1}{\sqrt{2\left(1\pm {a_{12}}/{a}\right)}}\left[\frac{1}{1-\bar{m}^2}-\frac{\bar{m}}{(1-\bar{m}^2)^{3/2}}\arccos \bar{m}\right].\label{NI}
\end{align}

\begin{figure}[ht]
\centering
\includegraphics[width=16cm]{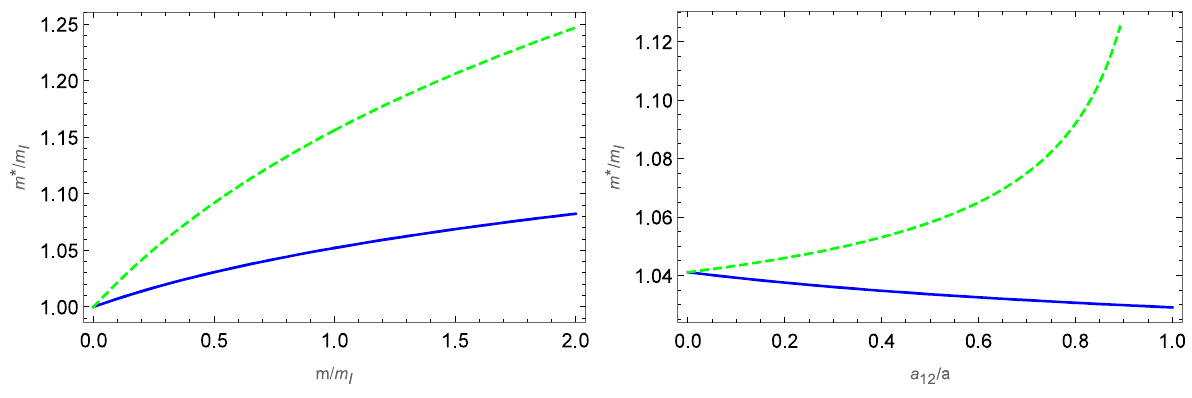}
\caption{The effective mass of polarons as a function of $m/m_I$ (a) and as a function of $a_{12}/a$ (b). The solid line represents $W^+$, while the dashed line represents $W^-$. The chosen parameters are consistent with Fig.\ref{Ema}.}\label{Wma}
\end{figure}

As displayed in Fig.\ref{Wma}(a), $W^-$ increases rapidly than $W^+$. In Fig.\ref{Wma}(b), we show the behaviors of $1+W^+$ and $1+W^-$ in the miscible region. We observe that the $W^-$ exhibits a sharp increase as $a_{12}/a$ approaches 1, while $W^+$ decreases slowly. These behaviors correspond to $I_W^{\pm}\sim (1\pm a_{12}/a)^{-1/2}$ in Eq.(\ref{WI}), in contrast to $I_{EC}^{\pm}\sim (1\pm a_{12}/a)^{1/2}$ as shown in Eq.(\ref{E+}).

The expressions of Eq.(\ref{em}), Eq.(\ref{E}), and Eq.(\ref{N}) clearly demonstrate that the characteristics of polarons are determined by the effective interaction. By examining the effective interaction Eq.(\ref{c1}) and Eq.(\ref{c2}) in the specific two cases mentioned above, we observe that a decrease in the effective interaction with the intercomponent scattering length, for example, leads to an increase in the ground state energy. This is because the effective interaction between the impurity and the Bogoliubov excitation causes a negative energy shift for the ground state energy of the polaron. Therefore, as the effective interaction decreases, the ground state energy increases. Conversely, an increase in the effective interaction with the scattering length between components leads to a decrease in the ground state energy. These phenomena correspond to $E_p^+$ and $E_p^-$, respectively.

\begin{figure}[ht]
\centering
\includegraphics[width=16cm]{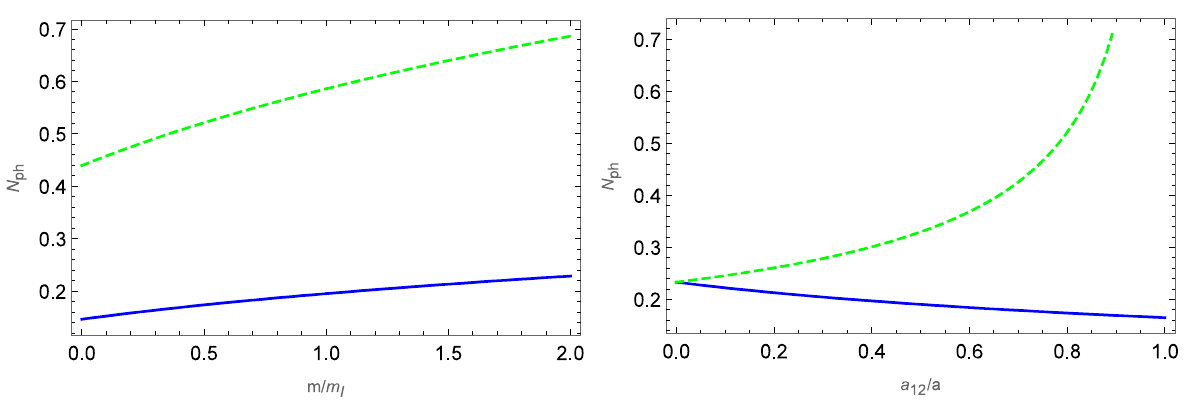}
\caption{ The average number of virtual phonons as a function of $m/m_I$ (a) and as a function of $a_{12}/a$ (b). The parameters are the same as Fig.\ref{Ema}. }\label{Nma}
\end{figure}

The results of Eq.(\ref{N}) are presented in Fig.\ref{Nma}, and they exhibit similar behavior to Fig.\ref{Wma}. For the contribution of ID couplings, the decrease in the average number of virtual phonons allows the impurity to be less tightly bound by the surrounding particles, resulting in a reduction in the effective mass. With fewer phonons, the impurity experiences less hindrance from the surrounding particles, leading to an increase in its kinetic energy. On the other hand, for the contribution of IS couplings, the increase in the average number of phonons hinders the movement of the impurity, leading to an increase in the effective mass of polarons and a decrease in its kinetic energy.

\subsection{Mass Imbalance}\label{r2}
\begin{figure}[ht]
\centering
\includegraphics[width=16cm]{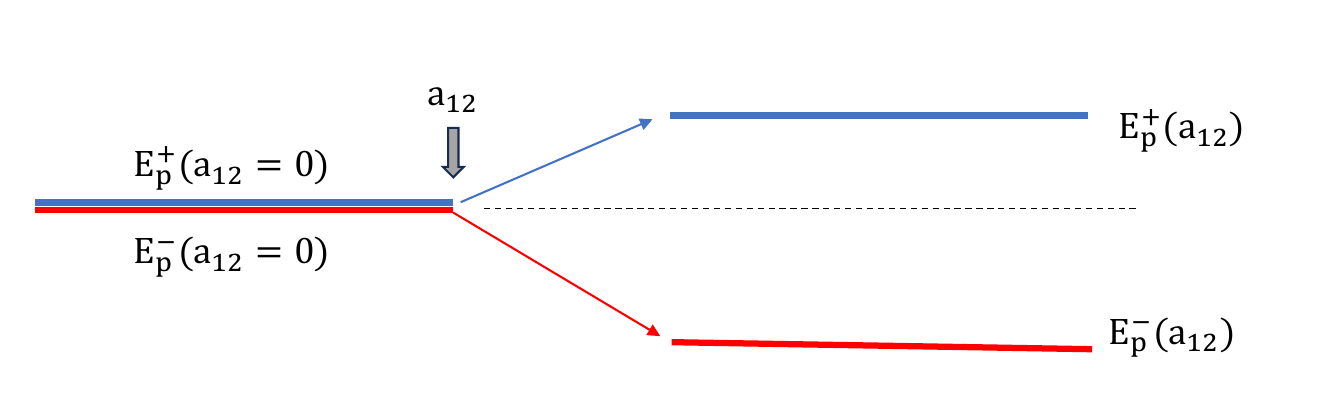}
\caption{The ground state energy shift of $E_p^+$ and $E_p^+$.  When the masses of the components are equal, the $E_p^+$ and $E_p^+$ are symmetric about the black dotted lines, but when the masses of the components are not equal, they are asymmetrical. }\label{e1e2}
\end{figure}

In this section, we consider the properties of Bose polaron in heteronuclear BECs, where two components have unequal masses. We focus on the contribution of mass imbalance to the polaron energy. We first discuss the case of homonuclear BECs. We expand $a_{12}/a$ as a small quantity in the ground state energy obtained from Eq.(\ref{EI}) and Eq.(\ref{EIS}) in the case of equal masses. Here, we have $\sqrt{\left(1\pm{a_{12}}/{a}\right)}\sim 1\pm {a_{12}}/{2a}$. It can be observed that the contribution resulting from ID and IS couplings is opposite and the absolute values of these contributions are equal. However, in the case of unequal-mass BECs, we demonstrate that the absolute values of these contributions are not equal. These discussions are depicted in Fig.\ref{e1e2}.

Next, we derive the properties of polarons caused by the mass imbalance of components. It is important to note that in the system of an impurity immersed in binary BECs, there are three mass ratios: $\bar{m}_1\equiv m_1/m_I$, $\bar{m}_2\equiv m_2/m_I$, and $\bar{m}_B\equiv m_1/m_2$. However, only two of these mass ratios are independent, meaning that if any two of the three ratios are known, the third one can be determined. Therefore, the energy of polarons depends on two independent variables of mass ratios. To clearly illustrate the structure induced by the mass imbalance, we set $a=a_{11}=a_{22}$, $\bar{m}_B<1$, and consider $a_{12}\ll a$. In this case, Eq.(\ref{ce}) is simplified to:
\begin{align}
  \bar{E}_{1}&=\bar{k}\sqrt{\bar{k}^2+2}\left[1+\left(\frac{a_{12}}{a}\right)^2\frac{M}{(\bar{k}^2+2)^2}\right]\label{BE1}\\
  \bar{E}_{2}&=\bar{m}_BE_{1}\label{BE2}
\end{align}
where $\bar{E}_{j}=E_{j}/(V_{11}n)$ and $M=\bar{m}_B(1+\bar{m}_B)/(1-\bar{m}_B)$. The effective interactions are given by:
\begin{align}
  \tilde{V}_{1I}^2&=\frac{\bar{k}}{\sqrt{\bar{k}^2+2}}V_{1I}^2\left[1+\frac{a_{12}}{a}\frac{2\bar{m}_B}{1-\bar{m}_B}\frac{1}{\bar{k}^2+2}\frac{V_{2I}}{V_{1I}}\right],\\
  \tilde{V}_{2I}^2&=\frac{\bar{k}}{\sqrt{\bar{k}^2+2}}V_{1I}^2\left[\left(\frac{V_{2I}}{V_{1I}}\right)^2-\frac{a_{12}}{a}\frac{2\bar{m}_B}{1-\bar{m}_B}\frac{1}{\bar{k}^2+2}\frac{V_{2I}}{V_{1I}}\right].\label{BEff}
\end{align}
By inserting Eq.(\ref{BE1})-(\ref{BEff}) into Eq.(\ref{E}) and ignoring terms of second order and higher in $a_{12}/a$, we obtain the expression for the polaron energy, $E_p=\sum_jV_{iI}N_{j,0}+\sum_{+,-}E_p^{\pm}$, where $E^+_p$ can be expressed as:
\begin{equation}
  E_p^+/E_{s1}= \frac{4}{\sqrt{3\pi^3}}\frac{(k_B a_{1I})^2}{(k_Ba)^{1/2}}(1+\bar{m}_1)^2\left(I^{+}_{E0}+\frac{a_{12}}{a}I^{+}_{E1}\right),\label{IDm}
\end{equation}
where $E_{sj}=2\pi\hbar^2 a n/m_j$. We obtain $I_{E0}^+=I^{+}_{EC}(a_+=2)$ , which is the result of Eq.(25) in Ref.~\cite{Huang2009}. It represents the ground state energy in the single-component BEC. The integral in the first-order term of $a_{12}/a$ for $E^+_p$ is given by:
\begin{align}
  I^{+}_{E1}(\bar{m}_1,\bar{m}_B)=\frac{k_Ba_{2I}}{k_Ba_{1I}}\frac{2\bar{m}_B(\bar{m}_B+\bar{m}_1)}{(1-\bar{m}_B)(1+\bar{m}_1)}\int_0^\infty\frac{\bar{k}^2{\rm d}\bar{k}}{(\bar{k}^2+2)^2+\bar{m}_1\bar{k}(\bar{k}^2+2)^{3/2}}.
  \end{align}
The result of this integral is given by:
  \begin{align}
  I^{+}_{E1}(\bar{m}_1,\bar{m}_B)=\frac{k_Ba_{2I}}{k_Ba_{1I}}\frac{\sqrt{2}\bar{m}_B(\bar{m}_B+\bar{m}_1)}{\bar{m}_1^2(1-\bar{m}_B)(1+\bar{m}_1)}\left(\frac{\arccos \bar{m}_1}{\sqrt{1-\bar{m}_1^2}}+\bar{m}_1-\frac{\pi}{2}\right).
\end{align}
We also obtain the expression of $E_p^-$,
\begin{equation}
   E_p^-/E_{s2}= \frac{4}{\sqrt{3\pi^3}}\frac{(k_B a_{2I})^2}{(k_Ba)^{1/2}}(1+\bar{m}_2)^2\left(I^{-}_{E0}+\frac{a_{12}}{a}I^{-}_{E1}\right).\label{ISm}
\end{equation}
Here, $I^-_{E0}=\mathcal{P}I^+_{E0}$, where the permutation operator $\mathcal{P}$ is defined as $\mathcal{P}f(m_1,m_2)=f(m_1 \rightarrow m_2,m_2 \rightarrow m_1)$. We show that it is a substitution of $\bar{m}_1$ in $I_{E0}^+$ with $\bar{m}_2$. The integral in first order term of $a_{12}/a$ for $E_p^-$ is as follows:
\begin{align}
  I^{-}_{E1}(\bar{m}_2,\bar{m}_B)=-\frac{k_Ba_{2I}}{k_Ba_{1I}}\frac{2\bar{m}_B^{-1}(\bar{m}_B^{-1}+\bar{m}_2)}{(\bar{m}_B^{-1}-1)(1+\bar{m}_2)}\int_0^\infty\frac{\bar{k}^2{\rm d}\bar{k}}{(\bar{k}^2+2)^2+\bar{m}_2\bar{k}(\bar{k}^2+2)^{3/2}}.
  \end{align}
The result of this integral is given by:
  \begin{align}
  I^{-}_{E1}(\bar{m}_2,\bar{m}_B)=-\frac{k_Ba_{2I}}{k_Ba_{1I}}\frac{\sqrt{2}\bar{m}_B^{-1}(\bar{m}_B^{-1}+\bar{m}_2)}{\bar{m}_2^2(\bar{m}_B^{-1}-1)(1+\bar{m}_2)}\left(\frac{\arccos \bar{m}_2}{\sqrt{1-\bar{m}_2^2}}+\bar{m}_2-\frac{\pi}{2}\right).
\end{align}

\begin{figure}[ht]
\centering
\includegraphics[width=10cm]{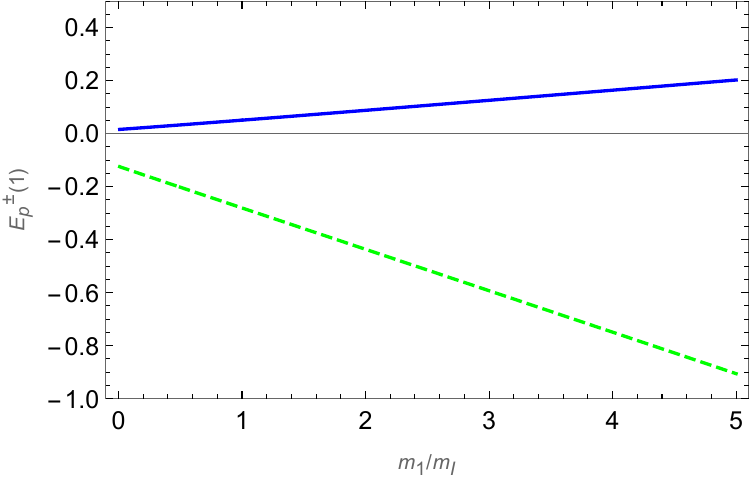}
\caption{Polaron energy induced by the intercomponent interaction $a_{12}$ with $m_1/m_2=0.5$ and $a_{12}/a=0.1$. $E_p^{\pm}(1)$ represents the last term of Eq.(\ref{IDm}) and Eq.(\ref{ISm}), respectively. The solid line represents $E_p^+(1)$ and the dashed line represents $E_p^+(1)$. The other parameters are consistent with Fig.\ref{Ema}.}\label{Em1ds}
\end{figure}

In Fig.\ref{Em1ds}, we show the energy shift, namely, the first-order term of $a_{12}/a$ in Eq.(\ref{IDm}) and Eq.(\ref{ISm}).
We find that there is a connection between the expressions of $E_p^{\pm}$. When components have no interaction ($a_{12}=0$), they have the same expressions. Once we introduce the intercomponent interaction, i.e.,  $a_{12}\ne 0$, $E^{\pm}_p(1)$ emerge. The connection of $E_p^+$ and $E_p^-$ is as follows:
\begin{align}
  E_p^-=\mathcal{P}E_p^+\label{cn}.
\end{align}

Similarly, we establish a relationship in terms of the effective mass and average number of phonons:
\begin{equation}
  W^-=\mathcal{P}W^+,\quad  N^-_{\rm ph}=\mathcal{P}N^+_{\rm ph}.\label{pmn}
\end{equation}
Detailed calculations are shown in Appendix \ref{A2}.

It is important to note that the connections Eq.(\ref{cn}) and Eq.(\ref{pmn}) we have shown are valid in a weak intercomponent coupling. We observe that the contribution resulting from ID and IS couplings can be connected through permutation operators, which provides a straightforward way to comprehend the behavior of Bose polaron in binary BECs. It remains uncertain whether these relations hold in other scenarios, such as strongly coupled multicomponent systems. We find that the contribution of IS coupling to the polaron energy tends to zero when approaching the miscible-immiscible region.

\section{Conclusion}\label{c}

In this study, we investigated the properties of polarons immersed in a two-component BEC. By performing a general Bogoliubov transformation, we derived the effective Hamiltonian to characterize Bose polarons, which includes two types of effective interactions: ID and IS couplings. Notably, we obtained a series of analytic results that characterize the properties of the polaron. We find that the contribution resulting from IS couplings to the ground energy decreases to zero near the miscible-immiscible boundary. Additionally, the effective mass of the polaron and the average number of virtual phonons become very large. The contribution of ID coupling to the polaron energy is similar to that of a single component. We also establish a link between the contribution of IS and ID coupling to the properties of polarons in homonuclear and heteronuclear BECs.

Finally, it is worth emphasizing that since we are considering weak coupling under the Fr\"{o}hlich framework, the theory described in this paper cannot accurately predict strong coupling when the coupling strength is close to the Feshbach resonance. We look forward to discussing strongly coupled models in future work, and we believe that richer phenomena such as many-body bound states will emerge. We anticipate extending this study to cases with coherent coupling and exploring other multiple-component systems, including the coupling of highly excited Rydberg states. Additionally, it would be interesting to investigate the properties of charged impurities or ion polarons in multiple-component BECs. Our results provide a solid foundation for understanding extensive polarons in two- or more-component BECs.

\begin{appendix}

\section{Effective interactions }\label{A1}

\begin{figure}[!htpb]
\centering
\includegraphics[width=16cm]{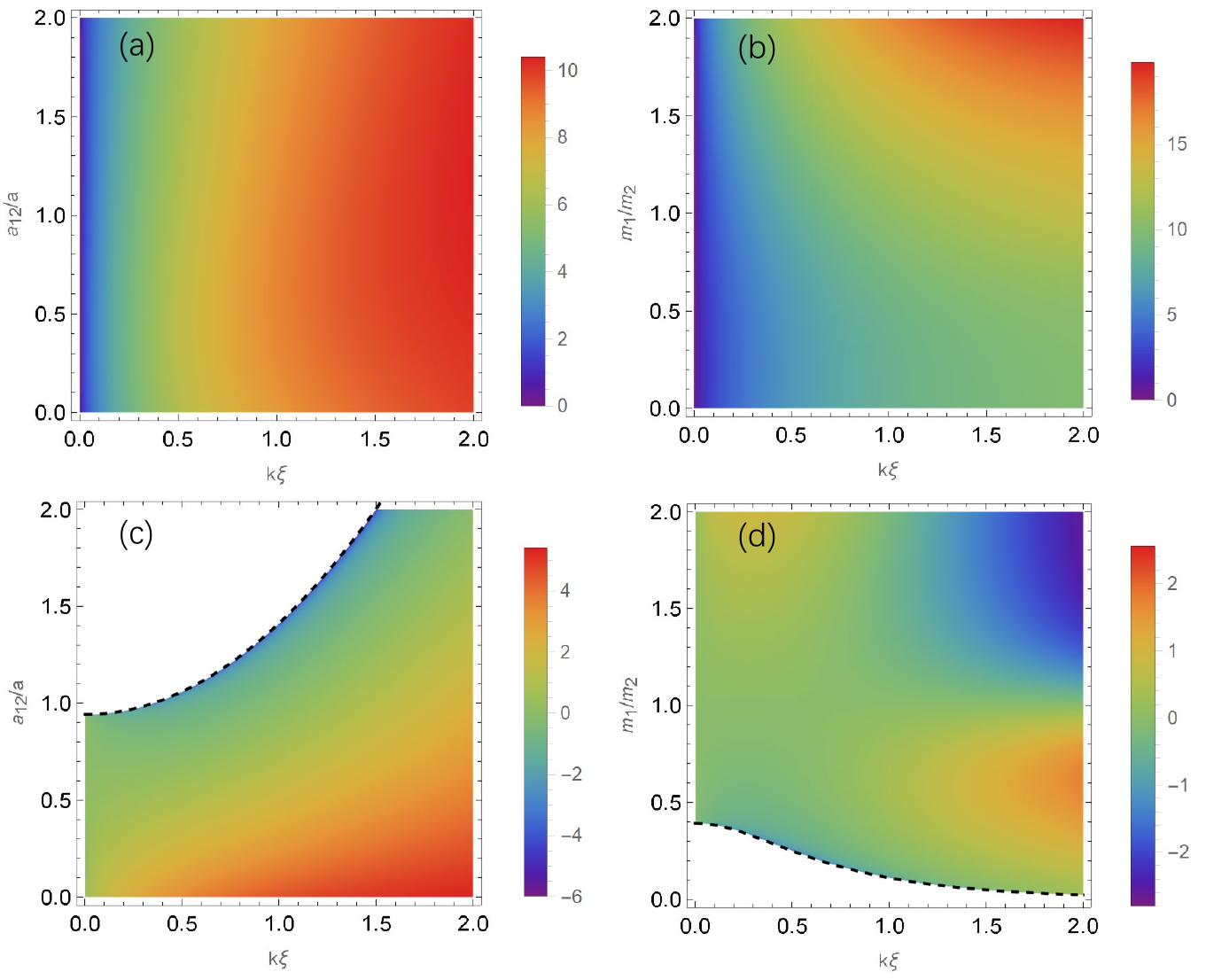}
\caption{(color online) Behaviors of effective interactions $\tilde{V}_{jI}/V_0$ with $V_0=2\pi\hbar^2a_{1I}/m_I$. We set $a=a_{11}=a_{22}$, $a_{1I}=a_{2I}$, and $n=n_{1,0}=n_{2,0}$. ID couplings are shown in (a) and (b), while IS couplings are shown in (c) and (d). In (a) and (c), the parameters are set as $m_1/m_2=1/2$ and $m_I/m_1=10$, while in (b) and (d), $a_{12}/a=0.9$ and $m_I/m_1=5$. The black dashed line in (c) and (d) represents the stability-unstability boundary. }\label{veff}
\end{figure}

In this appendix, we present the complete behavior of ID couplings and the IS couplings in Fig. \ref{veff}. As shown in Eq.(\ref{ev}), the behavior of the effective interaction is complex, unlike the monotonic behavior with intercomponent scattering length discussed in Fig. \ref{vc12} in the two special cases.
We illustrate the ID coupling in Fig. \ref{veff}(a) and Fig. \ref{veff}(b), and observe the ID coupling increases as $k$ increases. However, for a specific $k$, the ID coupling with $a_{12}$ is not monotonic.  Fig. \ref{veff}(c) and Fig. \ref{veff}(d) show the IS coupling. It is evident that, unlike the ID coupling, there is a distinct division where the blank and colored areas are separated by a dotted line, which corresponds to the unstable and stable regions, respectively. The dotted line corresponds to Eq.(\ref{sc}) which does not ensure that the lower excitation of Eq.(\ref{ce}) is real.

\section{Effective mass and the average number of virtual phonons in the case of unequal masses  }\label{A2}
In this appendix, we present calculations of the effective mass and the average number of virtual phonons in heteronuclear BECs, where $m_1\ne m_2$. Similar to the calculation process for the ground state energy, we only need to provide the contribution of ID couplings, $W^+$($N_{\rm ph}^+$), and $W^-$($N_{\rm ph}^-$) automatically emerges through the permutation of $m_1$ and $m_2$.

The effective mass of polarons and average number of phonons are given by $W=\sum_{\{+,-\}}W^{\pm}$ and $N_{\rm ph}=\sum_{\{+,-\}}N^{\pm}_{\rm ph}$. The expressions of $W^{\pm}$ and $N_{\rm ph}^{\pm}$ are as follows:
\begin{align}
  W^{\pm}&=\frac{8}{3\sqrt{3\pi^3}}\frac{(k_B a_{1I})^2}{(k_Ba)^{1/2}}\bar{m}_{1}(1+\bar{m}_{1})^2\left(I_{W0}^{\pm}+\frac{a_{12}}{a}I_{W1}^{\pm}\right),\label{IDW}\\
N^{\pm}_{\rm ph}&=\frac{2}{\sqrt{3\pi^3}}\frac{(k_B a_{1I})^2}{(k_Ba)^{1/2}}(1+\bar{m}_{1})^2\left(I^{\pm}_{N0}+\frac{a_{12}}{a}I_{N1}^{\pm}\right).\label{IDN}
\end{align}
Here,
\begin{align}
  I_{W0}^+&=I_W^+(a_+=2),\ I_{W0}^-=\mathcal{P}I_{W0}^{+},\\
  I^{+}_{N0}&=I_{N}^{+}(a_+=2),\ I^{-}_{N0}=\mathcal{P}I^{+}_{N0},
\end{align}
are results of Eq.(19) and Eq.(21) in Ref.~\cite{Huang2009}. This is the result of the effective mass of the polaron and the average number of phonons in the single-component BEC.
The integrals of $I_{W1}^{\pm}$ and $I^{\pm}_{N0}$ are as follows:
\begin{align}
  I_{W1}^+&=\frac{k_Ba_{2I}}{k_Ba_{1I}}\frac{2\bar{m}_B(\bar{m}_B+\bar{m}_1)}{(1-\bar{m}_B)(1+\bar{m}_1)}\int_0^\infty\frac{\bar{k}^2}{(\bar{k}^2+2)^{3/2}}\frac{{\rm d}\bar{k}}{(\sqrt{\bar{k}^2+2}+\bar{m}_1\bar{k})^3},\\
   I_{N1}^+&=\frac{k_Ba_{2I}}{k_Ba_{1I}}\frac{2\bar{m}_B(\bar{m}_B+\bar{m}_1)}{(1-\bar{m}_B)(1+\bar{m}_1)}\int_0^\infty\frac{\bar{k}}{(\bar{k}^2+2)^{3/2}}\frac{{\rm d}\bar{k}}{(\sqrt{\bar{k}^2+2}+\bar{m}_1\bar{k})^2}.
\end{align}
The results of the above integral are
\begin{align}
 I_{W1}^+&=\frac{1}{2\sqrt{2}\bar{m}_1^4}\frac{\bar{m}_B}{1-\bar{m}_B}\frac{\bar{m}_1+\bar{m}_B}{(1+\bar{m}_1)^2(1-\bar{m}_1)}\nonumber \\ &\times\left[3\pi+\bar{m}_1(5\bar{m}_1^2-3\bar{m}_1\pi-6)-(2\bar{m}_1^4-9\bar{m}_1^2+6)\frac{\arccos\bar{m}_1}{\sqrt{1-\bar{m}_1^2}}\right],\\
I_{N1}^+&=\frac{k_Ba_{2I}}{k_Ba_{1I}}\frac{\bar{m}_B(\bar{m}_B+\bar{m}_1)}{\sqrt{2}\bar{m}_1^3(1-\bar{m}_B)(1+\bar{m}_1)}\left[\frac{\bar{m}_1^2-2}{\sqrt{1-\bar{m}_1^2}}\arccos\bar{m}_1-2\bar{m}_1+\pi\right],\\
  I_{W1}^-&=\mathcal{P}I_{W1}^+,\quad I_{N1}^-=\mathcal{P}I_{N1}^+.
\end{align}
Thus we obtain the relation in Eq.(\ref{pmn}).
\end{appendix}

\begin{acknowledgments}
The authors are grateful for financial support from the National Natural Science Foundation of China (Grant No. 11975050).
    \end{acknowledgments}


\end{document}